\documentclass[aps,pre,preprint,superscriptaddress]{revtex4}
\usepackage{amsmath,amssymb,graphicx,color}
\usepackage{epstopdf}

\begin{document}
\title{Analysis of diffusion trajectories of anisotropic objects }

\author{Sunghan Roh}
\affiliation{Graduate School of Nanoscience and Technology,
Korea Advanced Institute of Science and Technology, Deajeon 305-701, Korea}

\author{Juyeon Yi}
\affiliation{Department of Physics, Pusan National University,
Busan 609-735, Korea}

\author{Yong Woon Kim}
\affiliation{Graduate School of Nanoscience and Technology,
Korea Advanced Institute of Science and Technology, Deajeon 305-701, Korea}


\begin{abstract}
We theoretically analyze diffusion trajectories of an anisotropic object moving on a
two dimensional space in the absence of an external field. In determining diffusion parameters associated with
the shape anisotropy, we devise a measure based on the gyration tensor, and obtain
its analytic expression exactly. Its efficiency and
statistical convergence are examined in comparison with the fourth cumulant of particle displacement.
We find that the estimation of diffusion constants based on the gyration measure is more efficient than analysis
adopting the fourth cumulant.

\end{abstract}


\maketitle

\section{Introduction}

Particles suspended in fluid move along random trajectories. Study on so called the Brownian motion
has provided fundamental understanding of macroscopic diffusion phenomena led by collective dynamics of microscopic constituents.
Theoretical framework to describe the Brownian motion
was proposed by Einstein~\cite{Einstein}, which is experimentally verified by
Perrin~\cite{PerrinExp}.
The Brownian trajectories reflect medium properties, shape and size of
particles. Hence, analyzing diffusive motions is a useful way to examine the physical properties of solution and colloid suspension, and its applications extend to the realm of chemistry and biology, let alone physics~\cite{Hanggi, Frey, Bressloff, Rice, Kampen}.

Although the original theory by Einstein describes translational motion in terms of diffusion constants,
theory was extended to rotational motion of an ellipsoid~\cite{Perrin1, Perrin2}, and an object with shape anisotropy has acquired considerable attention over decades for both in free space \cite{Kirkwood, Favro, Saffman, Prager, Lubensky, Sekimoto} and under an external potential field \cite{Yaliraki}.
There exist abundant examples of anisotropic objects such as
proteins and microtubles in nature, man-made colloidal particles~\cite{Snoeks, Dillen}, and nanotubes~\cite{Tsyboulski, Fakhri, Haghighi}.
It is worthwhile to mention a few diffusion characteristics of anisotropic particles. Since object more easily diffuses in the direction of its major principal axis, trajectory of an anisotropic particle becomes elongated along the preferred direction~\cite{Kirkwood}.
This asymmetric trajectories in turn allow us infer the shape of a particle, but it can only be observable within time interval
while orientation of the particle is preserved.  Also, the statistics of diffusion trajectory is not Gaussian having higher order cumulants which contain the information associated with shape~\cite{Prager, Lubensky, Sekimoto}.

It is only recent that these features could be experimentally characterized~\cite{Lubensky}. The state of art experiment was enabled by
single particle tracking technique~~\cite{Saxton},  and demonstrates a potential possibility to acquire shape information from diffusion
trajectories. Yet, the practical application is limited by a few factors. First is to realize high temporal resolution. If time resolution is longer than
orientation correlation time, diffusion trajectory becomes indiscernible from that of spherical or point particles in the laboratory frame, and information
related to shape anisotropy is totally erased. This issue would be partially solved by recent developments to enhance the resolution less than
orientational correlation time of membrane proteins of cells, which is typically order of $10 \sim 100 \mu$ \mbox{sec}~\cite{Fujiwara, Yeagle}.  More crucial is the fact that it is experimentally infeasible to identify initial angles of nanoscale objects, and one has to consider averaged quantity over an ensemble of randomly oriented particles.
In conventionally used second cumulant of displacement, terms associated with shape anisotropy vanish upon averaging over randomly distributed initial angles.
We thus needs to consider higher orders as relevant quantities to the trajectory analysis
in relation to the shape factor~\cite{Prager, Sekimoto, Lubensky}.
Since statistical convergences of higher order cumulants are usually poor, their statistical behaviors and errors are still to be examined.
Furthermore, natural question arises: What is the best higher order quantity to pursue efficient and reliable estimation of diffusion constants and shape anisotropy?

The purpose of this study is to propose a new measure, based on the gyration tensor, and suggest a systematic way to read
diffusion constants of anisotropic objects from particle position trajectories. Considering motions of anisotropic objects moving on a
two dimensional space in the absence of external force, we analyze the fourth order fluctuation of particle displacement both analytically
and numerically by Langevin dynamics simulation. In particular, we obtain the analytic expression of the gyration measure, which completely agrees with
Langevin dynamics simulation results. We compare the statistical convergence and finite sampling error of the gyration measure with previously suggested fourth moments
and find that analysis adopting gyration tensor is most efficient.
This paper is organized as follows. In Sec. II we introduce our system and its governing equations of motion, where we
illustrate an example of the diffusive trajectory, and discuss the properties of the second cumulant of particle position.
The higher order quantities are examined in Sec. III,  where we introduce the fourth order quantity obtained from the gyration tensor
and obtain its analytic expression exactly. In order to test the efficiency of those measures, we evaluate the standard errors of
the fourth cumulants and the gyration tensor based measure and explicate their errors in the estimation of diffusion constants in
Sec.IV.  {\color{blue} Finally in Sec. V, we discuss a possible scheme to implement least error measure and the properties of other fourth order measures.}   Summary will follow in Sec.VI.

\section{System}
\begin{figure}
\includegraphics[width=0.8\linewidth]{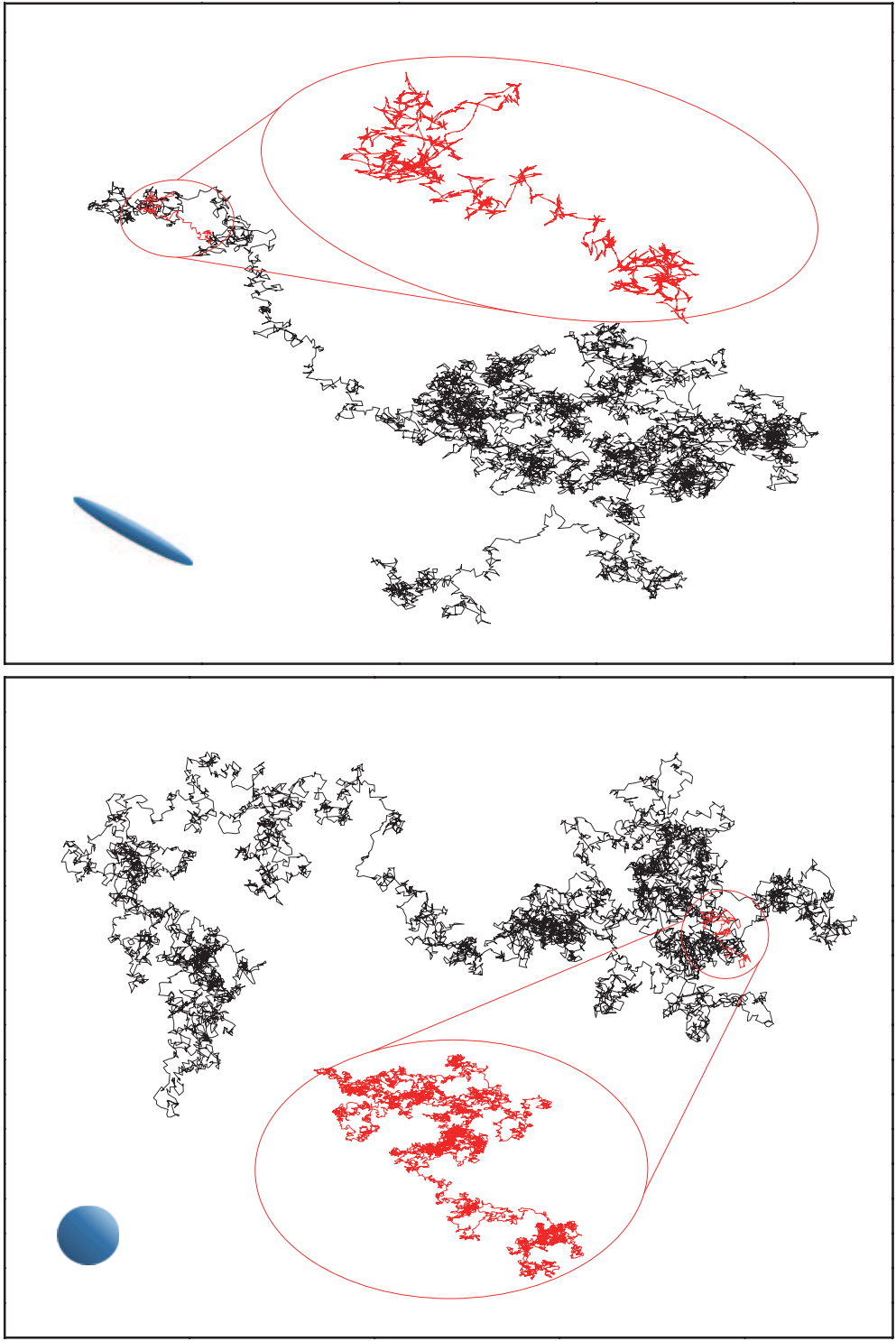}
\caption{Two dimensional diffusive motion of a highly anisotropic object with $D_{\perp}=0$~(the upper panel) in comparison with a spherical particle (the lower panel).
Trajectories are obtained by Langevin dynamic simulation, where total time elapse is $10^4/D_{\theta}$ with time resolution ($\Delta t$) chosen to be equal to the angular relaxation time $1/D_{\theta}$ for the whole trajectory. The local trajectories (see the blow ups) are acquired for finer time resolution
$\Delta t = 0.01/D_{\theta}$, and display distinctive features depending on object shape. }
\label{fig:system}
\end{figure}

We consider a Brownian particle of arbitrary shape, moving on a two dimensional space.
In the absence of external force, the translational motion of the Brownian particle can be described by the positional Langevin equation as
\begin{equation} \label{Eq:Position}
r_i \left( t \right) = \int_0^t \,d t' \xi_i \left( t' \right), \quad i=x, y,
\end{equation}
where $(r_x, r_y)$ denotes the center of mass coordinate, and the noise components $\xi_{x}$ and $\xi_{y}$ have zero means
and statistical properties given as
\begin{equation} \label{Eq:Correlation}
\left\langle \xi_i \left( t \right) \xi_j \left( t' \right) \right\rangle _{\theta}= 2D_{ij}(\theta) \delta \left( t-t'
\right).
\end{equation}
The noise amplitude is determined by diffusion matrix $D_{ij}(\theta)$.  Here we define the particle orientation coordinate $\theta$ to be an angle
between $x$ axis and one of eigenvectors of diffusion matrix $D_{ij}$, and $\langle \cdots \rangle_{\theta}$ denotes the noise average for
a fixed orientation $\theta$.
On the other hand, the equation of rotational motion of the particle is similarly given by
 \begin{equation} \label{Eq:Orientation}
\theta \left( t \right) = \int_0^t \,d t' \xi_\theta \left( t' \right),
\end{equation}
where $\xi_{\theta}$ is white noise with $\langle \xi_{\theta}(t)\rangle =0$ and $\langle \xi_{\theta}(t)\xi_{\theta}(t')\rangle = 2D_{\theta}\delta(t-t')$
with $D_{\theta}$ being the rotational diffusion constant.
Here we define the particle orientation coordinate $\theta$ to be an angle between $x$ axis and one of eigenvectors of diffusion matrix $D_{ij}$.
Note that $D_{ij}$ is a symmetric matrix which is diagonalizable to have eigenvalues, say,  $D_\parallel$ and $D_\perp$ (diffusion constants
along major and minor axis, respectively), and in laboratory coordinates, can be
represented as:
\begin{equation} \label{diffmat}
D_{ij} \left( \theta \right) = (1/2)[A\delta_{i,j}+S (\sigma_{z}\cos2\theta +\sigma_{x}\sin 2\theta)]
\end{equation}
with $A=(D_{\parallel}+D_{\perp})$ and $S=(D_{\parallel}-D_{\perp})$, and $\sigma_{\alpha}$ being the Pauli matrices.
Three diffusion constants, $A$, $S$ and $D_\theta$, fully characterize motion of the object, where nonzero $S$ indicates anisotropic motion,
resulting from shape anisotropy. In order to verify the analytic results to be derived shortly, and in order to explicitly evaluate
finite sampling errors, we employ the Langevin dynamics simulation in which position and orientations are updated by integrating
randomly generated noises, Eqs~(\ref{Eq:Position}) and~(\ref{Eq:Orientation}), satisfying respective fluctuation dissipation theorem. Time is discretized in the simulation with unit time step, say $\Delta \tau$. Length of $\Delta \tau$ should be much smaller than characteristic time scale of the phenomenon dealt in the simulation, which is the orientational relaxation time, $1/D_{\theta}$, for this case. We set $\Delta \tau = 0.01/D_{\theta}$ for generating trajectories presented in Figure.~1. Positions of the object are recorded at every time $\Delta t$, which is time resolution of the observation.

Figure~1 displays typical translational trajectories of a particle described by Eqs.~(\ref{Eq:Position}) and (\ref{Eq:Correlation}), where we compare the diffusion trajectory of an ellipsoid to that of a sphere.  The trajectories are obtained by choosing a time resolution equal to the orientation relaxation time,
$\Delta t =1/D_{\theta}$, and the trajectories in the marked areas are acquired by higher time resolution, $\Delta t = 0.01/D_{\theta}$. While the overall shape of diffusion trajectories of a sherical particle remains self similar upon changing the time resolution, the trajectory of the anisotropic particle
traced with the shorter time resolution appears different from the coarsely resolved trajectory, signaling
the correlation between angular and translational motion.

In characterizing stochastic trajectories, it is instructive to examine the moments of  a particle position at a given time. In the system of our interest, the first moment is identically zero due to the noise property. Using Eqs. (\ref{Eq:Position}) and (\ref{Eq:Correlation}), one finds the second moment,
\begin{eqnarray} \label{Eq:XSecondM}
\nonumber \left\langle r_x^2 \right\rangle_{\theta} &=& \int_0^t \,dt_1 \int_0^t \,d t_2
\left\langle \xi_x \left(t_1\right) \xi_x \left(t_2\right) \right\rangle_{\theta} \\
&=& A t + \frac{S}{4 D_\theta} \cos{2\theta} \left(1-e^{-4 D_\theta t}\right)~,
\end{eqnarray}
where the average is performed over ensemble with a fixed initial angle $\theta$. The second moment of $y$ coordinate of the particle trajectories can be obtained by replacing $S$ with $-S$. For an isotropic object~($S=0$), the second moment describes well the normal diffusion and the position dispersion is perfectly symmetric, as indicated by the aspect ratio, $\langle r^{2}_{y}\rangle_{\theta}/\langle r^{2}_{x}\rangle_{\theta} =1$. For $S\neq 0$, the diffusive behavior in the long time limit
($D_{\theta}t \gg 1$) is similar  to the normal diffusion with diffusion constant $A/2$, and the aspect ratio of trajectories is almost unity. If
observation time is short compared to the angular relaxation time ($D_{\theta}t \ll 1$), the diffusion follows the behavior, $\langle r_{x,y}^{2}\rangle _{\theta}\approx (A\pm S\cos2\theta) t$, dependent on the anisotropy factor $S$ and the initial angle, and leads to the aspect ratio deviated from unity.

\section{Fourth order measures}

In order to read $S$ and $D_{\theta}$ from the second moment given by Eq.~(\ref{Eq:XSecondM}), one needs to realize an ensemble of particles at a fixed initial angle,
which can possibly be done if one detects the initial angles of particles,
and collect a number of trajectories of particles which have the identical initial orientation.
However, it is difficult to resolve the orientations of microscopic objects in single particle tracking experiment~\cite{Sekimoto} so that experimentally available
are quantities averaged over randomly distributed initial angles.  Meanwhile as is apparent from Eq. (\ref{Eq:XSecondM}) the average of the second moment over initial angle $\theta$ nullifies the second term containing factors $S$ and $D_{\theta}$. Therefore, it is necessary in most practical
situations to examine higher moments in order to obtain the physical constants of objects from the initial angle averaged trajectories.
To the end, the authors of Ref. \cite{Lubensky} obtained the fourth cumulant of particle positions as
\begin{equation} \label{Eq:4thcumulant}
\langle C_4 \left( t \right)\rangle= \frac{3S^2}{16D^{2}_\theta}\left[4 D_{\theta} t + e^{-4 D_\theta t}-1\right]
\end{equation}
with $C_{4}(t)\equiv r_{x}^4 - 3\left\langle r_{x}^2 \right\rangle^2$. Here the average is also performed over initial angles,
and the diffusion constants $S$ and $D_{\theta}$ appear as relevant parameters to the fourth cumulant.

There certainly exist alternative ways to measure fluctuations in quartic order of the stochastic variable. For example, one can consider the fourth moment of translocation, $\langle r^{4}\rangle $. Following the calculation scheme in Appendix A,  we find,
\begin{equation} \label{Eq:M4}
\langle r^{4}\rangle = 8 A^2 t^2 + \frac{2S^2}{D_\theta} t - \frac{S^2}{2 D_{\theta}^2} \left( 1- e^{-4 D_{\theta} t} \right).
\end{equation}
Subtracting the first term given from the second moment of $r$ leads to a measure, $M_{4}\equiv r^{4}- 2\langle r^{2}\rangle^{2}$:\begin{equation}\label{m4}
\langle M_{4} \rangle= \frac{S^2}{2D^{2}_\theta} (4D_{\theta} t+ e^{-4 D_\theta t}-1 ).
\end{equation}
For short time measurements $\left( D_\theta t \ll 1 \right)$, $\langle M_{4}\rangle $ increase quadratically in time and in the long time measurement
$\left( D_\theta t \gg 1 \right)$, it converges to linear function of time. Such asymptotic behaviors are determined by $S$ and $D_{\theta}$.  This is also the case for $\langle C_{4}\rangle $ because of a relation $\langle C_{4}\rangle=3\langle M_{4}\rangle/8$.

Another fourth moment measure is gyration tensor, which is useful in identifying the spatial distribution of particles \cite{Solc,Theodorou,Arkin}.
The gyration tensor is constructed based on particle positions:
\begin{equation} \label{Eq:gyration}
G_N = \frac{1}{2N^2} \sum_{i,j=1}^{N}
\left(
\begin{array}{cc}
x^{2}_{ij} & x_{ij}y_{ij} \\
 x_{ij}y_{ij} & y_{ij}^{2}
\end{array}
\right),
\end{equation}
where $N$ is the total number of particles. Here distance between the $i$th and the $j$th particle are
denoted as $x_{ij}=r_{x,i}-r_{x,j}$ along the $x$-axis, and $y_{ij}=r_{y,i}-r_{y,j}$ along the $y$-axis.
The gyration tensor, Eq. (\ref{Eq:gyration}), has two non-negative eigenvalues, say $\lambda_1$ and $\lambda_2$ which act as shape identifiers and do not depend on the choice of coordinate.
Addition of eigenvalues, $\lambda_+ = \lambda_1 + \lambda_2$ gives the squared radius of gyration which corresponds to the size of particle distribution. The difference
of the eigenvalues,
$\lambda_- = \left| \lambda_1 - \lambda_2 \right|$, signifies the anisotropy of the distribution; $\lambda_{-}/\lambda_{+}$ vanishes for
an isotropic distribution, and becomes close to unity
if the distribution is highly anisotropic.

\begin{figure}
\includegraphics[width=0.9\linewidth]{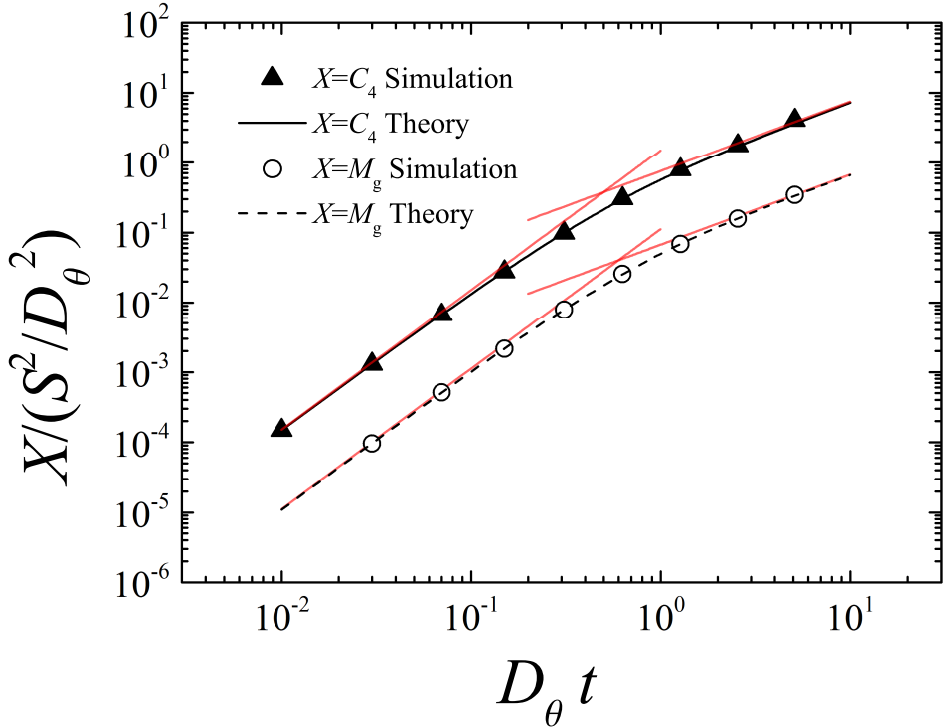}
\caption{Quartic order measures for analyzing diffusive motion of an anisotropic object ($S/A=1$) as a function of observation time $t$ in units of the
orientational relaxation time $D_{\theta}^{-1}$, where numerical results (points) from Langevin dynamic simulations are averaged
quantities acquired from $10^6$ simulations.
In comparison,  analytic expressions (lines), Eqs.~(\ref{Eq:4thcumulant}) and (\ref{mg}), are plotted,
showing an excellent agreement with numerical results. Straight lines are the asymptotic behaviors given in Eq.~(\ref{Eq:longshort}).}
\label{fig:system}
\end{figure}

For the present problem to analyze particle trajectories, we consider that $r_{x,i}$ and $r_{y,i}$ in Eq. (\ref{Eq:gyration}) represent the
position of a particle at discrete time $t_i$, that is, $r_x \left( t_i \right) = r_{x,i}$, and $r_{y}(t_{i})=r_{y,i}$. {\color{blue} For the fact that $\lambda_{-}$ quantifies
an anisotropy of distributions of $r_{\alpha,i}$'s,  one may take $\lambda_{-}^{2}$ as a quartic order measure. However, the average of $\lambda_{-}^{2}$ contains terms
independent of $S$ and $D_{\theta}$, which should be subtracted to define a relevant measure.
We introduce a {\it gyration} measure as $M_{g} = \lambda_{-}^2-f_A$ with $f_A$ given by \cite{notefA}:
\begin{equation} \label{Eq:fA}
f_A = \frac{2}{5} \left( \frac{1}{12} \left\langle r^2 \right\rangle^2 - \left\langle \lambda_+ \right\rangle^2 \right),
\end{equation}
where $\langle \lambda_{+}\rangle = At/3$. } 
After some lengthy but straightforward calculations, we  obtain exact expression for $M_{g}$ (see the details given in Appendix B) :
\begin{eqnarray}\label{mg}
\langle M_g \rangle &=& \frac{S^2}{1920 D_\theta^2}\frac{F(2D_{\theta}t)}{(D_{\theta}t)^{4}} \\ \nonumber
 F(x)&=&15 e^{-2x} ( 1 + x)^2 +15(x^2-1) -10 x^3+ 4 x^5,
\end{eqnarray}
which is again function of  $S$ and $D_{\theta}$, and acts as  another relevant measure for analyzing the diffusive motions of anisotropic objects.

The fourth order measures discussed so far entail factors related to the shape anisotropy in their respective forms.
In particular asymptotic behaviors, which can be summarized as
\begin{equation}\label{Eq:longshort}
\langle X \rangle D_{\theta}^{2}/S^{2}\approx \left\{
\begin{array}{lll}
&\alpha (D_\theta t)^{2}~,&D_{\theta} t  \ll 1 \\
&\beta D_{\theta} t~,  &D_{\theta} t \gg 1~,
\end{array}
\right .
\end{equation}
allow us to extract the diffusion constants of anisotropic object. Here the coefficients for $X=C_{4}, M_{4}, M_{g}$ are $\alpha = 3/2, 4, 1/9$ and $\beta = 3/4, 2, 1/15$, respectively. This long time and short time behaviors are displayed in Fig.~2, where we compare numerical results obtained from Langevin dynamics simulation with the analytic results. It can be seen that the two methods yield consistent results. This consistency could be obtained only
for sufficient repetitions of simulation ($10^6$ simulations were performed in producing Fig.~2). If the number of repeated simulations (or experiments)
is insufficient, the estimate of $X$ would be randomly deviated from the analytic results. Therefore, in practical estimation of the diffusion constants
from a finite number of measurements, the statistical convergence
of the measures and also finite sampling errors in estimating the diffusion constants come into question.

{\color {blue}
Before proceeding, let us make a remark on the effect of a marker position. In single particle tracking experiments diffusion trajectory of an object is obtained by tracing a point-like marker allocated on
the object. In our consideration,  the marker is assumed to be placed at the center of mobility. Even if the marker position is displaced from the center by $\ell$,
one can show by repeating the similar calculation that the long time behavior of  Eq.~(\ref{Eq:longshort}) remains unaffected, while only the short time behavior is modified:
The effect of the off-center distance on the short time behavior can be completely described by replacing $S$ with $S-D_{\theta}\ell^{2}$ in Eq.~(\ref{Eq:longshort}).
From a dimensional analysis, one thus finds that this correction would be negligible when when $L^2 (D_{\parallel}/D_{\perp}) \gg \ell^2$
with $L$ the size of the elongated object.
}

\section{Standard errors and finite sample bias }

In order to investigate the statistical convergence, we consider the variance of the measures,
\begin{equation}\label{relfluc}
\Delta X = \sqrt{\langle X^2 \rangle-\langle X\rangle ^2} ~.
\end{equation}
Although it is the very basic idea of statistics, we briefly mention the statistical meaning of $\Delta X$.
Suppose that one has a data set of $X_{i}$'s with $X_{i}$ obtained
from the $i$th measurement (or simulation), and the total number of data, $n$, is relatively large so that one can apply
the central limit theorem. Then, the bias of the finite sampling average $\sum_{i=1}^{n}X_{i}/n\equiv {\overline X}_{n}$ from
the true average $\langle X\rangle$ can be written as
\begin{equation}\label{standard error}
{\overline X}_{n}-\langle X\rangle \sim \Delta X/ \sqrt{n}~.
\end{equation}
From the above equation we find that the number of samplings required to reach the desired accuracy, $\epsilon^{*}=({\overline X}_{n}-\langle X\rangle)/\langle X\rangle$, amounts to
\begin{equation} \label{sample number}
\sqrt{n^{*}}\sim \Delta X/(\epsilon^{*}\langle X\rangle)~.
\end{equation}
Therefore, $\Delta X$ in relative to the average $\langle X\rangle$ quantifies the finite sampling error and also required number of measurements.
\begin{figure}
\includegraphics[width=0.9\linewidth]{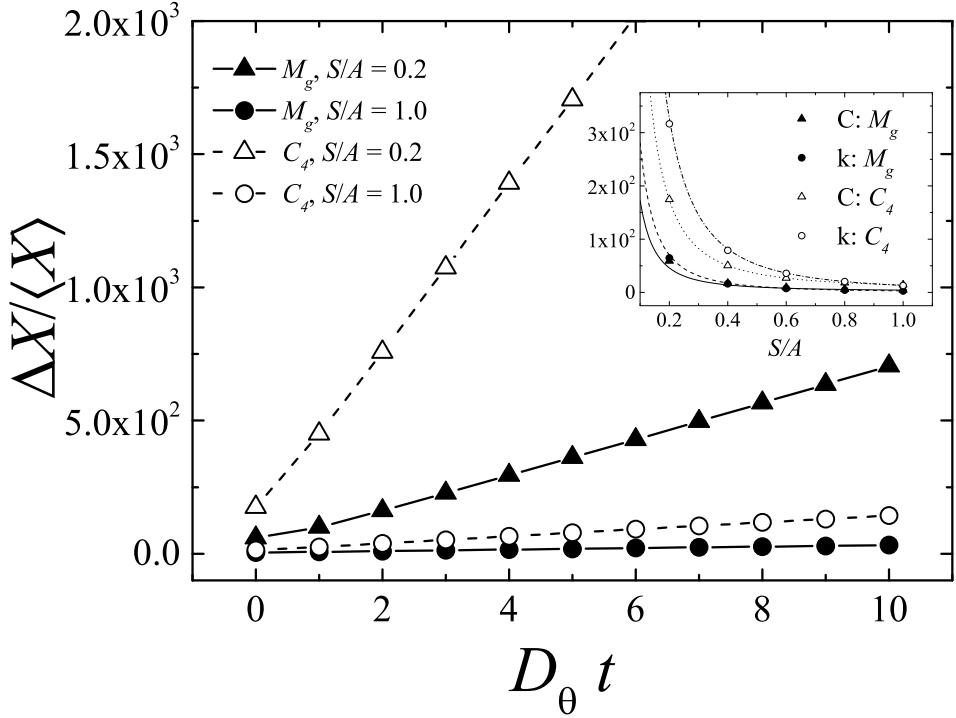}
\caption{Relative fluctuations, Eq.~(\ref{relfluc}), of the quartic order measures for various $S/A$ values as a function of $D_{\theta}t$,
where the average is taken over $10^6$ simulations. Relative fluctuation linearly increases in time, and it has an offset at $D_{\theta}t=0$,
showing the behavior given in Eq.(\ref{behavior}).
The inset shows the dependence of the offset and the slope on the shape factor $S/A$ and the choice of measure.
{\color{blue} Symbols represents values obtained from simulation while analytic expressions for $C$ for $M_g$(solid line), $k$ for $M_g$(dashed line), $C$ for $C_4$(dotted line), and $k$ for $C_4$(dash-dot line) are plotted for comparison.}
The fluctuations for $C_{4}$ are more significant than $M_{g}$ for a given $S/A$,
and for a given measure, it is more pronounced for smaller $S/A$.}
\label{Fig:ErrAndBehavior}
\end{figure}

In Fig.~\ref{Fig:ErrAndBehavior}, we present the relative fluctuation, $\Delta X/\langle X\rangle$, as a function of observation time. We observe that it increases linearly in time, and has finite offset at $t=0$, and hence
 \begin{equation}\label{behavior}
\Delta X /\langle X\rangle  \simeq k\left(D_\theta t\right)+C.
\end{equation}
Here the slop $k$ and the offset $C$ depend on the choice of measure and the shape
anisotropic factor $S/A$, as presented in the inset of Fig.~3.  {\color{blue} We also perform the detailed analysis of statistical errors of the gyration measure which indeed confirms Eq.~(\ref{behavior}) and allows us to analytically evaluate $k$ and $C$.}
Over the entire range of observation time, $\Delta X/\langle X\rangle$ for $X=C_{4}$ is
 bigger than that for the gyration measure $X=M_{g}$.
Moreover, $C_{4}$ increases more rapidly than $M_{g}$, and its statistical convergence in the long time limit should be much poorer than $M_{g}$.
This observation together with Eqs.~(\ref{standard error}) and (\ref{sample number}) implies that if one choose to use $C_{4}$ rather than the gyration measure,  estimation of $S$ and $D_{\theta}$ according to the limiting behaviors Eq.~(\ref{Eq:longshort}) from finite number of experiments can be more errorneous,
and requires more samplings.
In this respect it appears that more efficient analysis can be done by the gyration measure.
However,  because of $\Delta X/ \langle X\rangle$ linearly increasing in time, in using the gyration measure it is desirable to choose the observation time window, $t \lesssim t_{m}$ where $t_{m}$ is large for the long time behavior to set in, and yet not too large to bring in unwanted fluctuations.
It should also be noted that the relative fluctuation becomes more significant if we consider less anisotropic particles (smaller $S/A$).
{\color{blue} Note also in Appendix C that the magnitude of $k$ and $C$ are enhanced as $S/A$ decrease}.
This is because for small $S$ the measure itself becomes small, while $\langle X^{2}\rangle$ remains finite.
Therefore, it is expected that a large number of experiments should be repeated in order to probe weak anisotropy.

Let us now demonstrate more explicitly how the discussed behaviors of the variance are reflected in
the estimation of diffusion constants.  We evaluate finite sampling errors in $S$ and $D_{\theta}$,
and present the results in Fig.~4, where $S^{(n)}$ and $D_{\theta}^{(n)}$ represent values extracted
from finite sample average of $X$ of sample size $n$, ${\overline X}_{n}$, with $X=C_{4}$ and $X=M_{g}$.
In obtaining $S^{(n)}$, we measure the coefficient of $t^2$ of ${\overline X}_{n}$, using the short time behavior in Eq.~(\ref{Eq:longshort}).
In order to ensure $D_\theta \ll 1$ condition, the slope is evaluated at $t = 10^{-3}/D_\theta$ with unit time step $\Delta \tau = 10^{-4}/D_\theta$ .
In the upper panel of Fig.~4, we plot the bias, $|S^{(n)}-S|/S$ with $S$ being a true value, as a function of sample size.
As can be seen, the finite sample bias decreases with $n$, and follows well $1/\sqrt{n}$ behavior presented by the straight line.
For the smallest sample size $(n=10^{3})$, the error is roughly order of ten percent, and yet $C_{4}$ yields the error a few times larger
than $M_{g}$, which is due to its large fluctuation as previously stated.  The statistical fluctuations of $C_{4}$, more significant in the long
time limit, cause serious problem in the estimation of $D_{\theta}$. As shown in the lower panel of Fig.~4,  the bias ranges roughly from $1$ to
$10$, and moreover, {\color{blue} data points do not converge well along the line showing} $1/\sqrt{n}$ behavior due to its poor convergence. On the other hand, for the gyration measure,
although the error is large, it is order of magnitude less than the case of $C_{4}$, and more importantly, it can further be reduced by
increasing the sample size, as suggested in the bias well saturated into $1/\sqrt{n}$.

\begin{figure}
\includegraphics[width=0.9\linewidth]{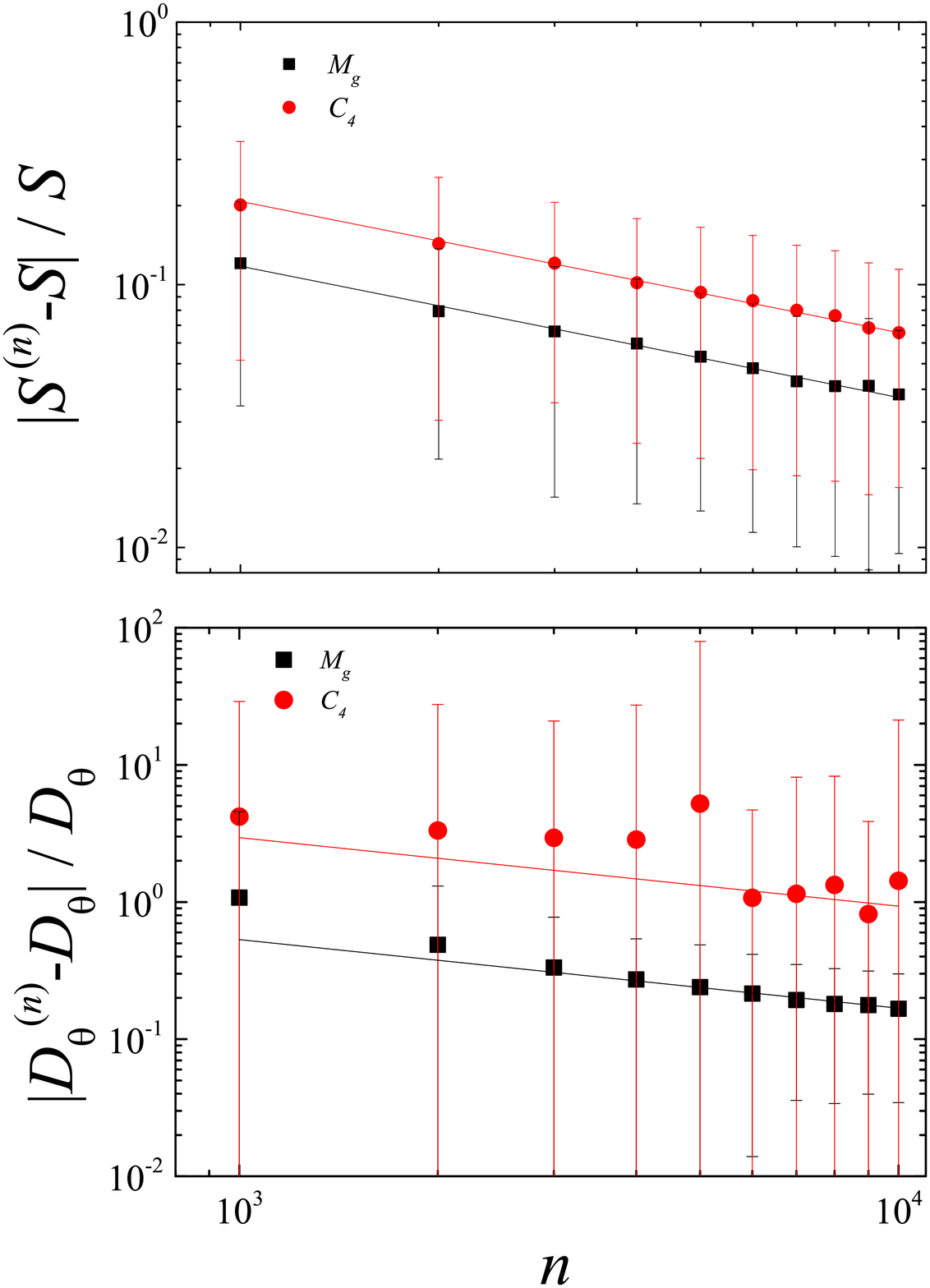}
\caption{Finite sampling errors as a function of sample size, where
$S^{(n)}$ and $D_{\theta}^{(n)}$ are the estimated values from measuring the asymptotic behaviors of ${\overline X}_{n}$.
The upper panel shows the bias of $S^{(n)}$ from its true value $S$. Both $C_{4}$ and $M_{g}$ result in the bias decreasing with the sample size
($1/\sqrt{n}$ lines are added for guides to the eye), but the bias is larger for $C_{4}$ in the entire range of sample size.
In determining $D_{\theta}^{(n)}$, we evaluate the slope of
${\overline X}_{n}$ at $t=D^{-1}_{\theta}$. The bias from $C_{4}$ is order of magnitude larger than $M_{g}$, and convergence to
$1/\sqrt{n}$ behavior is not yet achieved for the sample sizes in the plot range. We also choose the observation time $t=2/D_{\theta}$, although
{\color{blue} not} shown here, and find that the bias from $C_{4}$ becomes enlarged and does not show any consistent behavior as for the sampling size dependence.
}
\end{figure}

\section{Discussion}
{\color{blue}
In the preceding sections, we show that in analyzing trajectories of an anisotropic particle the gyration measure is more efficient  than
the fourth cumulant of displacement.  Before concluding our work, it may be instructive to discuss a formal procedure to implement the least error
measure and explain involved complications. One may write the generalized 4th order measure as,
\begin{eqnarray}
\mathcal{X} = \int_0^t \prod_{i=1}^{4}\,d t_i r_{\alpha_i,i} C_{\alpha_1 \alpha_2 \alpha_3 \alpha_4} \left( t, t_1, t_2, t_3, t_4 \right),
\end{eqnarray}
where $r_{\alpha_{i},i}$'s represent the position of particle along $\alpha_{i} = x, y$ direction at time $t_i$. The measure with least error should be invariant under permutation of $x$ and $y$ as well as interchange of time variables, and therefore, there are only three possibilities for $C_{\alpha_1 \alpha_2 \alpha_3 \alpha_4}$ to give
\begin{eqnarray} \label{Eq:GemeralMeasure}
\mathcal{X} &=&\prod_{i=1}^{4} \int_0^t \,d t_i [C_1 \left(x_1 x_2 x_3 x_4 + y_1 y_2 y_3 y_4\right) \\ \nonumber
&+& C_2 \left( x_1 y_2 y_3 y_4 + y_1 x_2 x_3 x_4\right) + C_3 \left( x_1 x_2 y_3 y_4 \right) ],
\end{eqnarray}
where $C_1 = C_{xxxx}$, $C_2=C_{xyyy}$, and $C_3=C_{xxyy}$. Arguments of $C_i$ are omitted and the sub-indices of $x$ and $y$ represent the indices of time variable. Note that ${\cal X}$
in general contains terms irrelevant to shape anisotropy.  With these anisotropy irrelevant terms
subtracted from ${\cal X}$, we obtain a measure, say, $X$.
Relative fluctuation determining the statistical error of a measure $X$ is given by $E[\{C\}] \equiv \Delta X/\langle X\rangle$ with $\Delta X$ defined in Eq.~(\ref{relfluc}). In order to acquire $X$ having least error, we need to build an equation for a function set $(C_1, C_2, C_3)$ which minimizes the functional $E[\{C\}]$. Although stating the problem may seem straightforward, there exist several hurdles to overcome.
One of those is without a function set of $C$'s specified, actual calculations are hardly doable. Besides, even if we choose a special set of $\{C\}$,
extremum condition for $E[\{C\}]$ leads to a complicated integral equation for $C$\cite{note}. We therefore cannot resort to this generalized scheme mentioned above.

Alternative way to find an efficient measure is to consider possible candidates of specific forms, and directly compare their statistical errors with one another,
which is the strategy chosen in this work. In addition to $C_{4}, M_{4}$ and $M_{g}$, we take other measures such as
\begin{eqnarray}
\nonumber
{\cal X}_{1}&=& \frac{1}{t} \int_0^t \,d t_1 x_1^4, \\ \nonumber
{\cal X}_{2}&=& \prod_{i=1}^2 \left[ \frac{1}{t} \int_0^t \,d t_i \right] \left( x_1^2 x_2^2 + y_1^2 y_2^2 \right), \\ \nonumber
{\cal X}_{3}&=&\prod_{i=1}^4 \left[ \frac{1}{t} \int_0^t \,d t_i \right] \left( x_1 x_2 x_3 x_4 + y_1 y_2 y_3 y_4 \right),
\end{eqnarray}
and examine their statistical properties. Here without calculation details, we only present obtained results that in the long time limit the errors of the measures listed above also linearly increase in time as
$\Delta X_{i}/\langle X_{i}\rangle = k_{i}D_{\theta}t$ with different coefficients, $k_{1}R^{2}=8\sqrt{38}/(3\sqrt{5})\approx 7.35 $,
$k_{2}R^{2}=2\sqrt{1838}/(3\sqrt{35})\approx 4.83$, and $k_{3}R^{2}=80/(9\sqrt{3})\approx 5.13$.
These values are larger than $k R^{2}=4\sqrt{31}/(3\sqrt{7})\approx 2.81$ for the gyration measure (see Appendix C), which substantiates that
the gyration measure outperforms not only the conventionally used fourth cumulant but also other measures considered above.
}

\section{Summary}

We considered the thermal motion of anisotropic objects in two dimensions, which is determined by three diffusion constants $D_{\parallel}$, $D_{\perp}$,
and $D_{\theta}$.  For the goal to determine the diffusion constants by analyzing diffusion trajectories, we investigated measures
of quartic order in particle position such as the fourth cumulant $\langle C_{4}\rangle$, the translocation measure $\langle M_{4}\rangle$,
and the gyration measure $\langle M_{g}\rangle $. The translocation measure is independent of initial orientation of a particle, and related to the
fourth cumulant as $\langle M_{4}\rangle = 8\langle C_{4}\rangle /3$. The measure $\langle M_{g}\rangle$ is obtained from eigenvalues of the gyration tensor, which identify the shape of particle trajectories.  We exactly obtained analytic expressions for the quartic order measures, and confirmed that numerical simulations give consistent results with the exact solution. It was found that the asymptotic behaviors of the measures are determined by the value of $S=D_{\parallel}-D_{\perp}$ and $D_{\theta}$, as given in Eq.~(\ref{Eq:longshort}). We further evaluated the variance of the measures
in order to test their statistical convergence. It was shown that the relative variance increases in observation time and has finite offset at
short time limit. The offset and the slope of the increment is small for $\left\langle M_{g}\right\rangle $, in comparison with $C_{4}$ or $M_{4}$, suggesting the efficiency of the gyration measure. We examined finite sampling error in the estimation of diffusion constants using $C_{4}$ and $M_{g}$,
and demonstrated explicitly that $M_{g}$ yielding smaller error outperforms the other measures.
We also showed that even using $M_{g}$, estimation of $D_{\theta}$ from the long time behavior can be severely biased due to the property of the variance increasing in measurement time, which is also the case for the evaluation of not only $D_{\theta}$ but also $S$ for a particle with weak anisotropy.

\section{acknowledgements}
This research was supported by Basic Science Research Program through the National Research Foundation of Korea(NRF) funded by the Ministry
of Education, Science and Technology(Grant No. NRF-2013R1A1A2013137).
S. R also acknowledges supported by NRF(National Research Foundation of Korea) Grant funded by the Korean Government(NRF-2013-Global Ph.D. Fellowship Program,
Grant No. NRF-2013H1A2A1033074).

\renewcommand{\thesection}{\Alph{section}}
\renewcommand{\theequation}{\Alph{section}.\arabic{equation}}
\setcounter{section}{1}

\appendix
\section{Evaluation of $M_{4}$}
We consider a displacement of a particle from an origin, $r = \sqrt{r_{x}^{2}+r_{y}^{2}}$, and evaluate its fourth moment,
\begin{equation}\label{appenm4}
\langle r^{4}\rangle = \langle r_{x}^{4}\rangle +2\langle r_{x}^{2}r_{y}^{2}\rangle + \langle r_{y}^{4}\rangle~.
\end{equation}
In this appendix, we let $\langle \cdots\rangle ^{x}$ to represent an average over a random variable $x$ which includes
$\xi_{x}, \xi_{y}, \xi_{\theta}$ and the initial angle.
According to Eq.~(\ref{Eq:Position}),
the positions can be expressed as the integrated noises, and the terms in Eq.~(\ref{appenm4}) can be written as
\[
\langle r_{\alpha_{1}}^{2}r_{\alpha_{2}}^{2}\rangle = \prod_{k=1}^{4}\int_{0}^{t} dt_{k}
\langle \xi_{\alpha_{1}}(t_{1})\xi_{\alpha_{1}}(t_{2})\xi_{\alpha_{2}}(t_{3})\xi_{\alpha_{2}}(t_{4})\rangle
\]
with $\alpha_{k}=x,y$.
Since the noises are Gaussian variables, one can apply Isserlis' theorem (or Wick's theorem) to write the average of quartic product of the nosies
in terms of the averages of paired noises:
\begin{eqnarray}
\langle \chi \rangle &\equiv & \langle  \xi_{\alpha_{1}}(t_{1})\xi_{\alpha_{1}}(t_{2})\xi_{\alpha_{2}}(t_{3})\xi_{\alpha_{2}}(t_{4}) \rangle
\\ \nonumber
&=&\langle \xi_{\alpha_{1},1}\xi_{\alpha_{1},2} \rangle \langle \xi_{\alpha_{2},3}\xi_{\alpha_{2},4}\rangle
+ 2\langle \xi_{\alpha_{1},1}\xi_{\alpha_{2},2}\rangle
\langle \xi_{\alpha_{1},3}\xi_{\alpha_{2},4}\rangle~,
\end{eqnarray}
where abbreviations $\xi_{\alpha_{i}}(t_{k})=\xi_{\alpha_{i},k}$ are used.  Using the property of the translational noise Eq.~(\ref{Eq:Correlation}),  we obtain
\begin{eqnarray}
\langle \chi \rangle^{\xi_{x},\xi_{y}} &=& {\cal D}_{ij}(\theta_{1},\theta_{3})\delta (t_{1}-t_{2})\delta(t_{3}-t_{4}) \\ \nonumber
{\cal D}_{ij}(\theta_{1},\theta_{3})&\equiv &4D_{ii}(\theta_{1})D_{jj}(\theta_{3})+8D_{ij}(\theta_{1})D_{ij}(\theta_{3})~,
\end{eqnarray}
where $\theta_{i} \equiv \theta(t_{i})$. Summing up the results obtained so far, and using the diffusion matrix components $D_{ij}$ given
in Eq.~(\ref{diffmat}), we reach
\begin{equation}\label{appen2}
\langle r^{4}\rangle ^{\xi_{x},\xi_{y}} = \int_{0}^{t}dt' \int_{0}^{t} dt'' [8A^{2}+4S^{2}\cos2\Theta(t',t'')]
\end{equation}
with $\Theta(t',t'')=\theta(t')-\theta(t'')$.
We are now left with average over angular noise and initial angle. This angular average process should be performed on the cosine term in the above equation.
For the Gaussian property of $\xi_{\theta}$, we have the following identity:
\begin{equation}
\langle e^{in[\theta(t)-\theta(t')]}\rangle ^{\xi_{\theta}}=e^{-n^{2}D_{\theta}|t-t'|}.
\end{equation}
Plugging this relation for $n=2$ into Eq.~(\ref{appen2}), integrating with respect to time leads to the fourth moment of translocation, Eq.~(\ref{Eq:M4}).

\section{Evaluation of $M_{g}$}
Let us first evaluate the average of $\lambda_{+}$ which is the sum of eigenvalues of $G_{N}$ in Eq.~(\ref{Eq:gyration}):
\begin{equation}\label{lamp}
\lambda_+ = \frac{1}{2N^2} \sum_{i, j = 1}^N \left[ \left( r_{x,i} - r_{x,j} \right)^2 + \left( r_{y,i} - r_{y,j} \right)^2 \right].
\end{equation}
We take $r_{\alpha, i}$ to represent the position of particle along $\alpha =x,y$ direction at time $t_{i}$. The averages of the components appearing in Eq.~(\ref{lamp}) are given by Eq. (\ref{Eq:XSecondM}) to lead
\begin{equation}
\langle \lambda_+ \rangle = \frac{A}{N^2} \sum_{i, j = 1}^N | t_i - t_j |.
\end{equation}
In the continuum limit of time, we can replace the summation with an integration as
\begin{equation}\label{conti}
N^{-1}\sum_{i=1}^{N} \approx t^{-1}\int_{0}^{t}dt'~,
\end{equation}
and obtain
\begin{equation}\label{lamp2}
\lim_{\Delta t \rightarrow 0}\langle \lambda_+ \rangle = \frac{1}{t^2} \int_0^t \,d t' \int_0^t \,d t'' A \left| t' - t'' \right| = \frac{1}{3} A t.
\end{equation}

The subtraction of the eigenvalues yields a lengthy expression of $\lambda_{-}^{2}$ given in terms of
combinations of quartic multiples of particle positions at different times. We give a short hand writing of the expression,
\begin{equation}\label{Eq:LambdaMinusSquared}
\lambda_{-}^2 = \frac{1}{N^4} \sum_{\{i_k\}= 1}^{N} [r^{(4)}_{x}+r^{(2)}_{x}r^{(2)}_{y}+(x\leftrightarrow y)],
\end{equation}
where the summation denotes a multiple summation, $\sum_{\{i_{k}\}} \equiv \sum_{i_{1}, i_{2},i_{3}, i_{4}}$, and $\left( x \leftrightarrow y \right)$ indicates terms produced by permuting $x$ and $y$ of the terms given in the square braket. The summand, $r^{(4)}_{x}$ is given only in terms of
$x$ coordinate:
\[
r^{(4)}_{x}= r_{x, i_1}^2 r_{x, i_2}^2 - 2 r_{x, i_1}^2 r_{x, i_2} r_{x, i_3} + r_{x, i_1} r_{x, i_2} r_{x, i_3} r_{x, i_4}~,
\]
and the mixing term $r^{(2)}_{x}r^{(2)}_{y}$ contains factors quadratic in both $x$ and $y$ coordinate:
\[
\begin{array}{lll}
r^{(2)}_{x}r^{(2)}_{y}
&=&2 r_{x, i_1}^2 r_{y, i_2} r_{y, i_3} + r_{x, i_1} r_{y, i_2} r_{x, i_3} r_{y, i_4}- r_{x, i_1}^2 r_{y, i_2}^2   \\
&+& 2 r_{x, i_1} r_{y, i_1} r_{x, i_2} r_{y, 2}- 4 r_{x, i_1} r_{y, i_1} r_{x, i_2} r_{y, i_3}~.
\end{array}
\]

Analytic expression for gyration measure are obtained in the continuum time limit where the approximation in Eq.~(\ref{conti}) can be applied.
We define $\left\langle {\cal M}_g \right\rangle = \lim_{\Delta t \rightarrow 0} \left\langle \lambda_{-}^2 \right\rangle$, and re-express it in concise form as
\begin{equation}
\left\langle {\cal M}_g \right\rangle = \sum_{\{\alpha_{k}\}} C_{\{\alpha\}}\left\langle \prod_{k=1}^4 \frac{1}{t}
 \int_{0}^{t}  d t_{k} r_{\alpha_k}(t_{k}) \right\rangle, \label{Eq:AveragedMg}
\end{equation}
where $\alpha_{k}=x,y$. Summation $\sum_{\{\alpha_{k}\}}$ runs over all the combinations appearing in Eq.~(\ref{Eq:LambdaMinusSquared}),
and the corresponding coefficient is given by $C_{\{\alpha\}}$. In obtaining $\langle {\cal M}_{g}\rangle$,
we first evaluate the average of quartic moment of position variables:
\begin{equation}
\left\langle \prod_{k=1}^4 r_{\alpha_k}(t_k) \right\rangle =
\left\langle \prod_{k=1}^4 \int_0^{t_{k}} d t'_{k} \xi_{\alpha_k}(t'_k)  \right\rangle~. \label{Eq:MgSpatialProd}
\end{equation}
Upon using the Isserlis' theorem, the average of quartic moment of noises can be written as
\begin{equation}
\nonumber \left\langle \prod_{k=1}^4 \xi_{\alpha_k}(t'_{k}) \right\rangle =
\langle 1,2\rangle \langle 3,4\rangle+\langle 1,3\rangle \langle 2,4\rangle+\langle 1,4\rangle \langle 2,3\rangle
\end{equation}
where $\langle k,\ell\rangle \equiv \left\langle \xi_{\alpha_k}(t'_k) \xi_{\alpha_\ell}(t'_{\ell}) \right\rangle$. Upon averaging over the translational noises,  we have
\[
\langle k,\ell\rangle ^{\xi_{x},\xi_{y}}=2 D_{\alpha_k \alpha_\ell} \left( \theta (t'_{k}) \right) \delta \left( t'_{k} - t'_{\ell} \right),
\]
and the time integration of these quantities in Eq.~(\ref{Eq:MgSpatialProd}) can be done, for which we define
\[
\phi_{k,\ell}\equiv 2 \int_{0}^{\textrm{min}(t_{k},t_{\ell})}dt'_{k}D_{\alpha_{k},\alpha_{\ell}}(\theta(t'_{k})).
\]
Here $\textrm{min} \left( t_{k}, t_{\ell} \right)$ is smaller value between $t_{k}$ and $t_{\ell}$.
Then, the average of quartic moment of spatial variables is given by
\[
\left\langle \prod_{k=1}^4 r_{\alpha_k}(t_k) \right\rangle = \langle \phi_{1,2}\phi_{3,4}+\phi_{1,3}\phi_{2,4}+\phi_{1,4}\phi_{2,3}\rangle^{\xi_{\theta},\theta_{0}}~,
\]
where the average over the angular noise can be derived through a relation for the Guassian property of $\xi_{\theta}$:
\[
\langle e^{in\Delta\theta(t)\pm im \Delta\theta(t')}\rangle^{\xi_{\theta}} = e^{-D_{\theta}[n^{2}t+m^{2}t'\pm 2\textrm{min}(t,t')mn ]}~
\]
with $\Delta \theta(t)\equiv \theta(t)-\theta_{0}$.
This gives the average of the fourth moment of position at different times in Eq.~(\ref{Eq:MgSpatialProd}) for a given set of $\alpha_{k}$'s.
Then ${\cal M}_{g}$ in Eq.~(\ref{Eq:AveragedMg}) can be evaluated by performing the integrations and summing over all the sets of $\alpha_{k}$'s
corresponding to the terms in Eq.~(\ref{Eq:LambdaMinusSquared}).  This process is lengthy but straightforward to reach
\begin{eqnarray}\label{calmg}
\nonumber \left\langle {\cal M}_g \right\rangle &=& \frac{4}{45}(At)^{2}+ \frac{1}{1920}\frac{S^2}{D^{6}_\theta t^{4}} [ 15 e^{-4 D_\theta t} (1+2 D_\theta t)^2-15 \\
&+&60 (D_\theta t)^{2}- 80 (D_\theta t)^3 + 128(D_\theta t)^5 ].
\end{eqnarray}
Defining $M_{g}$ to eliminate the shape irrelevant term, that is, $M_{g}={\cal M}_{g}-f_A$ with $f_A = 4(At)^{2}/45$,  we obtain Eq. (\ref{mg}).

\section{Error coefficients of the gyration measure}
{\color{blue}
The statistical error of the gyration measure is evaluated, which leads to explicit expression for $k$ and $C$ in the long and short time regime, respectively. In calculating the variance, $\Delta X$ for the gyration measure $X=M_{g}$, we use an identity,
\[
\Delta M_{g} = \sqrt{\langle M_{g}^{2}\rangle - \langle M_{g}\rangle^{2}}=
\sqrt{\langle {\cal M}_{g}^{2}\rangle - \langle {\cal M}_{g}\rangle^{2}}~.
\]
Given $\left\langle \mathcal{M}_g \right\rangle$ by Eq.(\ref{calmg}), we are left with evaluating $\left\langle \mathcal{M}_g^2 \right\rangle$. In long time regime, we only keep terms of highest order in time $t$ and obtain
\begin{eqnarray}
\nonumber
\left\langle \mathcal{M}^2 \right\rangle = 128 A^4 (3 L_1^4 -12 L_1^2 L_2 + 8 L_2^2 + 8 L_1 L_3 - 8 L_4  \\
\nonumber
+ 2 L_1^2 O_2 - 4 L_2 O_2 + O_2^2 +2O_4 )
+{\cal O}(t^{3}).
\end{eqnarray}
Here symbols $L_i$ and $O_i$ are defined below through an auxiliary variable, $p_j= \int_0^{t_j} \,d t' \zeta(t')$ with $\zeta(t)$ being
a Gaussian stochastic variable which satisfies $\left\langle \zeta(t') \right\rangle=0$ and $\left\langle \zeta(t) \zeta(t') \right\rangle = \delta(t-t')$:
\begin{eqnarray}\label{ol}
 L_1 &=& \prod_{i=1}^{2}I_{i}\left\langle p_1 p_2 \right\rangle = \frac{1}{3} t \\
\nonumber L_2 &=& \prod_{i=1}^{3}I_{i}\left\langle p_1 p_2 \right\rangle \left\langle p_2 p_3 \right\rangle = \frac{2}{15} t^2\\
\nonumber L_3 &=& \prod_{i=1}^{4}I_{i}\left\langle p_1 p_2 \right\rangle \left\langle p_2 p_3 \right\rangle \left\langle p_3 p_4 \right\rangle = \frac{17}{315} t^3 \\
\nonumber L_4 &=& \prod_{i=1}^{5}I_{i}\left\langle p_1 p_2 \right\rangle \left\langle p_2 p_3 \right\rangle \left\langle p_3 p_4 \right\rangle \left\langle p_4 p_5 \right\rangle = \frac{62}{2835} t^4 \\
\nonumber O_1 &=& I_{1} \left\langle p_1 p_1 \right\rangle = \frac{1}{2} t \\
\nonumber O_2 &=&   \prod_{i=1}^{2}I_{i}\left\langle p_1 p_2 \right\rangle \left\langle p_2 p_1 \right\rangle = \frac{1}{6} t^2 \\
\nonumber O_3 &=& \prod_{i=1}^{3}I_{i}\left\langle p_1 p_2 \right\rangle \left\langle p_2 p_3 \right\rangle \left\langle p_3 p_1 \right\rangle = \frac{1}{15} t^3 \\
\nonumber O_4 &=& \prod_{i=1}^{4}I_{i}\left\langle p_1 p_2 \right\rangle \left\langle p_2 p_3 \right\rangle \left\langle p_3 p_4 \right\rangle \left\langle p_4 p_1 \right\rangle = \frac{17}{630} t^4,
\end{eqnarray}
where an integral operation is defined as $I_{i}=\int_{0}^{t}dt_{i}/t$.
One can find that the relative variance or the statistical error is given by
\begin{equation}
\Delta M_{g}/\langle M_{g}\rangle
\approx \frac{4}{3} \sqrt{\frac{31}{7}} \frac{A^2}{S^2} D_{\theta}t,
\end{equation}
which confirms the behavior  of Eq.~(\ref{behavior}) linearly increasing in time and  correspondingly gives the slope $k$,
\begin{equation}\label{k}
k = \frac{4}{3} \sqrt{\frac{31}{7}} \frac{1}{R^2}
\end{equation}
with $R \equiv S/A$.

Next, let us evaluate the offset of $\Delta X/\langle X\rangle$, $C$, in short time limit. Considering that orientation is almost fixed during short time interval, we find after some tedious algebra that $\left\langle \mathcal{M}^2 \right\rangle$ is expanded as:
\begin{equation}\label{shortcalmg}
\left\langle \mathcal{M}^2 \right\rangle \simeq  128 A^4 \langle {\cal M}^{2}_{g}\rangle_{A}
+128 A^2 S^2  \langle {\cal M}^{2}_{g}\rangle_{AS}
+16 S^{4}\langle {\cal M}^{2}_{g}\rangle_{S}
\end{equation}
with the coefficients given by
\begin{eqnarray}
\nonumber
\langle {\cal M}^{2}_{g}\rangle_{A}&\equiv &3 L_1^4 -12 L_1^2 L_2 + 8 L_2^2 + 8 L_1 L_3 - 8 L_4 \\ \nonumber
&+& 2 L_1^2 O_2 - 4 L_2 O_2 + O_2^2 +2 O_4 \\ \nonumber
\langle {\cal M}^{2}_{g}\rangle_{AS} &\equiv& 9L_1^4 - 30 L_1^2 L_2 + 8 L_2^2 +28 L_1 L_3 - 16 L_4 \\ \nonumber
&-&6 L_1^3 O_1 + 16 L_1 L_2 O_1 - 12 L_3 O_1 + L_1^2 O_1^2 \\ \nonumber
&-&2 L_2 O_1^2 + L_1^2 O_2 - 2 L_1 O_1O_2 + O_1^2O_2  \\ \nonumber
&-& 4L_1O_3 + 4O_1O_3 + 4O_4 \\
\nonumber
\langle {\cal M}^{2}_{g}\rangle_{S}&\equiv & 9L_1^4 - 24L_1^2 L_2 + 16L_2^2 -12L_1^3O_1 + 16L_1L_2O_1 \\ \nonumber
 &+&10L_1^2O_1^2 - 8L_2O_1^2 - 4L_1O_1^3  + O_1^4 + 12L_1^2O_2 \\ \nonumber
 &-&16L_2O_2 -8L_1O_1O_2 + 4O_1^2O_2 + 4O_2^2~.
\end{eqnarray}
Inserting equations in (\ref{ol}) into Eq.~(\ref{shortcalmg}), we obtain
\[
\langle {\cal M}_{g}^{2}\rangle = \frac{608}{14175} A^4t^4 + \frac{2608}{14175} A^2 S^2t^4 + \frac{1}{25}S^4t^4~,
\]
and further using that $\langle M_{g}\rangle \approx S^{2}t^{2}/9$ in the short time limit, as given in Eq.~(\ref{Eq:longshort}), we finally reach
\begin{eqnarray}\label{c}
\nonumber C &=& \lim_{t\rightarrow 0} \Delta M_{g}/\langle M_{g}\rangle \\
&=& \frac{2}{5} \sqrt{\frac{124 + 582 R^2 + 98 R^4}{7}} \frac{1}{R^2}.
\end{eqnarray}
The analytic expressions of $k$ and $C$, respectively given in Eq.~(\ref{k}) and Eq.~(\ref{c}), are consistent with the simulation result shown in inset of Fig. 3.
}


\begin{thebibliography}{10}
\bibitem{notefA} There are various ways to choose $f_A$ as long as $f_A = 4(At)^2/45$ is satisfied. However, we recommend the form given in Eq.~(\ref{Eq:fA}) since it does not introduce terms associated with marker position displacement $l$, in long time regime.
\end{thebibliography}
\end{document}